\documentclass[12pt]{article}
\usepackage{amsfonts}
\usepackage{amssymb}
\usepackage{amsmath}
\usepackage{amsthm}
\usepackage{graphicx}

\begin{document}

\title{\bf Introduction to teleparallel gravities}

\author{\Large Alexey Golovnev${}^{a,b}$\\
${}^{a}${\it Asia Pacific Center for Theoretical Physics,}\\ 
{\it 67 Cheongam-ro, Nam-gu, Pohang 37673, Republic of Korea}\\
{\small alexey.golovnev@apctp.org}\\
${}^{b}${\it Faculty of Physics, St. Petersburg State University,}\\ 
{\it Ulyanovskaya ul., d. 1, Saint Petersburg 198504, Russia}\\
{\small agolovnev@yandex.ru}}
\date{}

\maketitle

\begin{abstract}

We give a pedagogical introduction into the field of (modified) teleparallel theories of gravity. Our presentation is fairly self-contained. In particular, we carefully explain the basic principles of metric-affine approaches to gravity. 

This contribution is based on our talk "Teleparallel gravity, its modifications, and the local Lorentz invariance" at the 9th Mathematical Physics Meeting: School and Conference on Modern Mathematical Physics in Belgrade, September 2017.

\end{abstract}

\section{Introduction}

The  great success of general relativity in describing experimental data does clearly show that we are on the right track, and our geometric theory of gravity works extremely well. On experimental side the only reservations come from cosmology. 

Even though we have a perfectly good model also there ($\Lambda CDM$), it is very unpleasant to realise that we don't have any reasonable idea as to the nature of some 95\% of the energy budget of the Universe \cite{Planck}. Numerous searches for new physics beyond the Standard Model of elementary particles are yet futile.

Can we do better via some (infrared) modifications of gravity? Very probably, the answer could be in positive. It seems to be not an easy endeavor, but nevertheless the one which is worth trying.

In order to have an interesting model, we need to start from a deep understanding of gravity. Simply changing the Newtonian limit for incorporation of, say, flat rotation curves in galaxies is way too easy. MOND is an archetypal example \cite{MOND}. Embedding it into a wider picture is very arbitrary and hardly was ever successfull per se \cite{Natasha}.

Of course, modifications to the full general relativity can also be done in a large number of different ways. However even the most straightforward consistency tests prove to restrict the freedom a lot. A very good example is massive gravity for which even the most basic task of avoiding the Boulware-Deser ghost turned out a highly non-trivial problem \cite{Claudia} with interesting mathematical structures behind \cite{GS1, GS2}. 

Even very simple extensions such as (metric) $f(R)$ models lead to a new degree of freedom due to the higher derivative nature of any available curvature invariants; and the problem of screening the extra degrees of freedom from low energy phenomenology emerges \cite{scr}. One notable exception from this rule is  Palatini $f(R)$ gravity which is however subject to its own problems \cite{palfr}.

We must conclude that the search for a viable modification to the theory of gravitational interactions is anything but a simple task, and very often the models acquire their own deep academic interest even irrespective of possible phenomenological applications. And despite a large number of reasonable ideas on the market, we cannot claim a definitive success of any particular approach to cosmology-motivated modification of general relativity.

From this vantage point, it would be natural to consider models of teleparallel type. Indeed, teleparallel gravity is just an equivalent (modulo global topological issues of parallelisability) description of general relativity in terms of different geometry. However, if one attempts at modifying the teleparallel Lagrangian then the resulting modified gravity models from simple modifications would generically be different from those which can be obtained by analogous procedures applied to general relativity in the standard formulation. Therefore it is an independent direction of research which is definitely worth trying.

It is precisely this viewpoint which we will take regarding modified teleparallel gravities in this contribution. In Section 2 we explain the essence of metric-affine approaches to gravity. In Section 3 we describe the teleparallel equivalent of general relativity in its classical standing with vanishing spin connection, and also present some possible extensions. In Section 4 we discuss covariantised versions and related problems. Finally, in Section 5 we conclude.

\section{Metric-affine preliminaries}

In principle, a metric and an affine connection are two different geometrical structures on a manifold. The former defines lenghts and distances, and also gives the natural measure $\sqrt{-g}\cdot d^4x$, while the latter provides us with the rule of parallel transport. Apparently, we don't need to think of the metric structure in order to have some choice of the horizontal section in the tangent bundle.

\subsection{Affine connection}

Leaving an invariant discussion for textbooks on differential geometry, we will use the plain language of components for the purposes of this elementary introduction. The connection coefficients $\Gamma$ are used to tell us how the components of a vector are changed under an infinitesimal parallel transport, $\delta x^{\mu}$ in terms of coordinates: $\delta A^{\nu}=-\Gamma^{\nu}_{\mu\alpha}A^{\alpha}\delta x^{\mu}$, or analogously for 1-forms $\delta A_{\nu}=\Gamma^{\alpha}_{\mu\nu}A_{\alpha}\delta x^{\mu}$. The same coefficients with different signs are used in these rules in order to ensure that scalar quantities such as $A_{\mu}B^{\mu}$ are not changed under the parallel transport.

One can treat the parallelly transported vector as being the same vector at a different point. Covariant derivatives are then simply defined by subtracting this trivial change above from the actual change of vector field components:
\begin{equation}
\label{covvec}
\bigtriangledown_{\mu}A^{\nu}\equiv\partial_{\mu}A^{\nu}+\Gamma^{\nu}_{\mu\alpha}A^{\alpha},
\end{equation}
and
\begin{equation}
\label{covform}
\bigtriangledown_{\mu}A_{\nu}\equiv\partial_{\mu}A_{\nu}-\Gamma^{\alpha}_{\mu\nu}A_{\alpha}
\end{equation}
for differential 1-forms. Note that different conventions can be encountered at this point. Ours is that the index which corresponds to the derivative is the left one among the lower indices of the connection coefficients.

The rules (\ref{covvec}) and (\ref{covform}) were given in one particular coordinate system. One can however deduce the transformation law for $\Gamma$ demanding that the covariant derivatives be tensors. It does not depend on any particular choice of the affine connection, and a simple calculation from any textbook on general relativity gives
\begin{equation}
\label{transcon}
\Gamma^{\kappa}_{\mu\nu}=
\frac{\partial x^{\kappa}}{\partial {x^{\prime}}^{\rho}}
\left(\frac{\partial {x^{\prime}}^{\alpha}}{\partial x^{\mu}}
\frac{\partial {x^{\prime}}^{\beta}}{\partial x^{\nu}}{\Gamma^{\prime}}^{\rho}_{\alpha\beta}+
\frac{\partial^2 {x^{\prime}}^{\rho}}{\partial x^{\mu} \partial x^{\nu}}\right).
\end{equation}

Now we see that symmetric connections, $\Gamma^{\alpha}_{\mu\nu}=\Gamma^{\alpha}_{\nu\mu}$, are somewhat distinguished from the viewpoint of the equivalence principle since they can be set to zero at any single point by a mere coordinate transformation. 
On the other hand, the antisymmetric part is a tensor
\begin{equation}
\label{torsion}
T^{\alpha}_{\hphantom{\alpha}\mu\nu}=\Gamma^{\alpha}_{\mu\nu}-\Gamma^{\alpha}_{\nu\mu}
\end{equation}
which is known under the name of torsion. It will be very important for us in what follows.

\subsubsection{Geodesic lines}

Given an affine connection, one can generalise the notion of a straight line. 
Indeed, let us call a line $x^{\mu}(\tau)$ geodesic if an only if its tangent vector $e^{\mu}(\tau)\equiv\frac{dx^{\mu}(\tau)}{d\tau}$ remains tangent when parallelly transported along the line. 

In other words, the tangent vector $e^{\mu}(\tau)$ is covariantly constant along the line which means that we demand
$\delta\frac{dx^{\mu}}{d\tau}=-\Gamma^{\mu}_{\nu\alpha}\frac{dx^{\alpha}}{d\tau}\delta x^{\nu}=-\Gamma^{\mu}_{\nu\alpha}\frac{dx^{\alpha}}{d\tau}\frac{dx^{\nu}}{d\tau}\delta\tau$  under an infinitesimal change of the parameter $\tau$. It gives the geodesic equation
\begin{equation}
\label{geodesic}
\frac{d^2x^{\mu}}{d\tau^2}+\Gamma^{\mu}_{\nu\alpha}\frac{dx^{\alpha}}{d\tau}\frac{dx^{\nu}}{d\tau}=0.
\end{equation}

This equation (\ref{geodesic}) is invariant under affine changes of $\tau$. Under a general non-linear reparametrisation, this equation becomes more complicated.
So, there is a preferred class of choices for the parameterisation of the line, the class of affine parameters.  It stems from demanding that $e^{\mu}(\tau)$ is strictly preserved when being parallelly transported along a geodesic line, not only up to a scalar factor, and allows one to define the null infinity in general relativity such that a null geodesic is infinitely long if its affine parameter reaches an infinite value.

\subsection{Curvature tensor}

Of course, if one simply introduces curvilinear coordinates in a Euclidean space, connection coefficients already become non-trivial.
One may ask how to distinguish between a real curvature and a (locally) flat space in curvilinear coordinates. In Riemannian geometry the answer is very simple (in general we will only need to add torsion and non-metricity to this discussion, see below). 

It is difficult to compare two vectors at a distance from each other, but it is fairly easy to do so if they are at the same point in the space. If one parallelly transports a vector along a closed contour, then the resulting vector always coincides with the initial one if the space is flat. Therefore, if after such a procedure a vector has changed, it is a clear indication of non-trivial geometry (the converse is also true but not so elementary, even for rigorous understanding of what it means).

Let us perform a parallel transport of a vector $\xi^{\mu}$ along an arbitrarily chosen closed infinitesimal contour $\mathcal C$. After one revolution we have
$$\delta\xi^{\mu}=\oint\limits_{\mathcal C}d\tau\cdot\frac{d \xi^{\mu}}{d\tau}=-\oint\limits_{\mathcal C}d\tau\cdot\Gamma^{\mu}_{\nu\alpha}\xi^{\alpha}\cdot\frac{dx^{\nu}}{d\tau}$$
where the vector function $\xi^{\mu}(x(\tau))$ at any point of the contour is determined by the parallel transport equation $\frac{d}{d\tau}\xi^{\mu}=-\Gamma^{\mu}_{\nu\alpha}\xi^{\alpha}\frac{dx^{\nu}}{d\tau}$ with an arbitary initial value at an arbitrary initial point on $\mathcal C$.  Moreover, we can imagine that the vector is parallelly transported  to everywhere in a small neighbourhood of the contour. Well, it is a little tricky given the non-uniqueness we want to establish, but at least we can definitely see an obstacle for the geometry to be trivial.

Let us assume for simplicity that the origin of the coordinate system, i.e. $x^{\mu}=0$, is chosen inside the contour and Taylor expand the vector field components and the connection coefficients around the origin.
Since at the lowest order the integral is proportional to $\oint dx^{\mu}=0$, we look at the first order correction:
$$\delta\xi^{\mu}=-\oint\limits_{\mathcal C}d\tau\left(\vphantom{\int}\Gamma^{\mu}_{\nu\alpha}\cdot(\partial_{\rho}\xi^{\alpha})x^{\rho}
+(\partial_{\rho}\Gamma^{\mu}_{\nu\alpha})x^{\rho}\cdot\xi^{\alpha}+{\mathcal O}(x^2)\right)\frac{dx^{\nu}}{d\tau}$$
where the only $x$-dependence is explicit one with all functions being taken at $x^{\mu}=0$.

In the first term we use again the parallel transport equation for the vector field  $\partial_{\rho}\xi^{\alpha}=-\Gamma^{\alpha}_{\rho\sigma}\xi^{\sigma}$  to obtain
$$\delta\xi^{\mu}=-\oint\limits_{\mathcal C}d\tau\left(\vphantom{\int}-\Gamma^{\mu}_{\nu\alpha}\Gamma^{\alpha}_{\rho\sigma}\xi^{\sigma}x^{\rho}
+(\partial_{\rho}\Gamma^{\mu}_{\nu\sigma})\xi^{\sigma}x^{\rho}+{\mathcal O}(x^2)\right)\frac{dx^{\nu}}{d\tau}.$$
Omitting the ${\mathcal O}(x^2)$-corrections, the integral is proportional to
$$\oint x^{\rho}\frac{dx^{\nu}}{d\tau}d\tau=-\oint x^{\nu}\frac{dx^{\rho}}{d\tau}d\tau=\int\int dx^{\rho}\wedge dx^{\nu},$$
the infinitesimal antisymmetric area element (encircled by the contour $\mathcal C$) which we will denote by $S^{\rho\nu}=-S^{\nu\rho}$. 

The antisymmetrised coefficient in front of $S^{\rho\nu}$ must also be a tensor, what can of course be checked directly by using equation (\ref{transcon}), and finally we get
\begin{equation*}
\delta\xi^{\mu}=\oint\limits_{\mathcal C}d\tau{\dot \xi}^{\mu}=-\frac12 R^{\mu}_{\phantom{\mu}\sigma\rho\nu}\xi^{\sigma}S^{\rho\nu}
\end{equation*}
where the curvature tensor is defined as
\begin{equation}
\label{curvature}
R^{\mu}_{\phantom{\mu}\sigma\rho\nu}=\partial_{\rho}\Gamma^{\mu}_{\nu\sigma}-\partial_{\nu}\Gamma^{\mu}_{\rho\sigma}
+\Gamma^{\mu}_{\rho\alpha}\Gamma^{\alpha}_{\nu\sigma}
-\Gamma^{\mu}_{\nu\alpha}\Gamma^{\alpha}_{\rho\sigma}.
\end{equation}
By definition, it is always antisymmetric in the last two indices
\begin{equation}
\label{ascurv}
R^{\mu}_{\phantom{\mu}\sigma\rho\nu}=-R^{\mu}_{\phantom{\mu}\sigma\nu\rho},
\end{equation}
and this is the only symmetry property of the curvature tensor for a generic connection.

Of course, we could also transport a one-form instead of a vector around the contour. It yields the same tensor:
\begin{equation*}
\delta\zeta_{\mu}=\oint\limits_{\mathcal C}d\tau{\dot \zeta}_{\mu}=\frac12 R^{\sigma}_{\phantom{\sigma}\mu\rho\nu}\zeta_{\sigma}S^{\rho\nu},
\end{equation*}
 where the change of sign ensures again that the scalar quantities remain unchanged.
 
\subsubsection{Commutator of covariant derivatives}

 There is also another way to introduce curvature. Normally, the partial derivatives commute. However, it is not true of the covariant ones. Let us compute the commutator:
\begin{multline*}
\left[\bigtriangledown_{\mu}\ ,\ \bigtriangledown_{\nu}\right]\xi^{\alpha}=
\bigtriangledown_{\mu}\left(\partial_{\nu}\xi^{\alpha}+\Gamma^{\alpha}_{\nu\beta}\xi^{\beta}\right)
-\bigtriangledown_{\nu}\left(\partial_{\mu}\xi^{\alpha}+\Gamma^{\alpha}_{\mu\beta}\xi^{\beta}\right)=\\
=\partial_{\mu}\left(\partial_{\nu}\xi^{\alpha}+\Gamma^{\alpha}_{\nu\beta}\xi^{\beta}\right)+
\Gamma^{\alpha}_{\mu\rho}\left(\partial_{\nu}\xi^{\rho}+\Gamma^{\rho}_{\nu\beta}\xi^{\beta}\right)-
\Gamma^{\rho}_{\mu\nu}\left(\partial_{\rho}\xi^{\alpha}+\Gamma^{\alpha}_{\rho\beta}\xi^{\beta}\right)-\\
-\partial_{\nu}\left(\partial_{\mu}\xi^{\alpha}+\Gamma^{\alpha}_{\mu\beta}\xi^{\beta}\right)-
\Gamma^{\alpha}_{\nu\rho}\left(\partial_{\mu}\xi^{\rho}+\Gamma^{\rho}_{\mu\beta}\xi^{\beta}\right)+
\Gamma^{\rho}_{\nu\mu}\left(\partial_{\rho}\xi^{\alpha}+\Gamma^{\alpha}_{\rho\beta}\xi^{\beta}\right)=\\
=R^{\alpha}_{\phantom{\alpha}\beta\mu\nu}\xi^{\beta}-T^{\rho}_{\hphantom{\rho}\mu\nu}\bigtriangledown_{\rho}\xi^{\alpha}
\end{multline*}
where $T^{\rho}_{\hphantom{\rho}\mu\nu}=\Gamma^{\rho}_{\mu\nu}-\Gamma^{\rho}_{\nu\mu}$
is the torsion tensor (\ref{torsion}). 

Therefore, we have
\begin{equation}
\label{commvec}
\left[\bigtriangledown_{\mu}\ ,\ \bigtriangledown_{\nu}\right]\xi^{\alpha}=R^{\alpha}_{\phantom{\alpha}\beta\mu\nu}\xi^{\beta}-T^{\rho}_{\hphantom{\rho}\mu\nu}\bigtriangledown_{\rho}\xi^{\alpha},
\end{equation}
or analogously for a 1-form:
\begin{equation}
\label{commform}
\left[\bigtriangledown_{\mu}\ ,\ \bigtriangledown_{\nu}\right]\zeta_{\alpha}
=-R^{\beta}_{\phantom{\beta}\alpha\mu\nu}\zeta_{\beta}-T^{\rho}_{\hphantom{\rho}\mu\nu}\bigtriangledown_{\rho}\zeta_{\alpha}.
\end{equation}

\subsection{Torsion and non-metricity}

In Riemannian geometry, the curvature tensor is a precise diagnostic tool to see if the geometry is (locally) trivial. For general metric-affine models it is no longer the case, and we need more information. Of course, one obvious indicator of non-trivial geometry is the torsion tensor (\ref{torsion})
$$T^{\alpha}_{\hphantom{\alpha}\mu\nu}=\Gamma^{\alpha}_{\mu\nu}-\Gamma^{\alpha}_{\nu\mu}.$$

One more tool is easy to discover if we recall that in Riemannian geometry the metric tensor is covariantly constant, $\bigtriangledown_{\alpha}g_{\mu\nu}=0$. Let us therefore introduce another tensorial quantity, the non-metricity tensor
\begin{equation}
\label{nonmetr}
Q_{\mu\alpha\beta}=\bigtriangledown_{\mu}g_{\alpha\beta}
\end{equation} 
which also signals non-trivial metric-affine geometry.

Now we finally  have all necessary ingredients for working with arbitrary connections in the sense that the postulated equation (\ref{nonmetr}),
$\partial_{\mu}g_{\alpha\beta}=\Gamma^{\rho}_{\mu\alpha}g_{\rho\beta}+
\Gamma^{\rho}_{\mu\beta}g_{\alpha\rho}+Q_{\mu\alpha\beta}$, together with a prescribed torsion
fixes the affine connection uniquely.

Indeed, following the standard derivation of the Levi-Civita connection, we compute the combination $\partial_{\mu}g_{\nu\beta}+\partial_{\nu}g_{\mu\beta}-\partial_{\beta}g_{\mu\nu}$ and use it to determine the components of $\Gamma$, this time keeping track of its antisymmetric part which is $T$ and of $Q\neq 0$. The result is
\begin{multline}
\label{arbconn}
\Gamma^{\alpha}_{\mu\nu}=\frac12 g^{\alpha\beta}\left(\partial_{\mu}g_{\nu\beta}+\partial_{\nu}g_{\mu\beta}-\partial_{\beta}g_{\mu\nu}\right)\\
+\frac12\left(T^{\alpha}_{\hphantom{\alpha}\mu\nu}+T^{\hphantom{\nu}\alpha}_{\nu\hphantom{\alpha}\mu}+T^{\hphantom{\mu}\alpha}_{\mu\hphantom{\alpha}\nu}\right)-\frac12\left(Q^{\hphantom{\mu\nu}\alpha}_{\mu\nu}+Q^{\hphantom{\nu\mu}\alpha}_{\nu\mu}-Q^{\alpha}_{\hphantom{\alpha}\mu\nu}\right).
\end{multline}

We see that possible connections differ from each other by contributions of torsion and non-metricity. As it should  have been, difference of two connections is always a tensor.

\subsubsection{Symmetric connections}

In case of symmetric connections we put torsion to zero, and the curvature tensor has further symmetry properties. A simple inspection of the definition (\ref{curvature}) readily shows that for symmetric $\Gamma$ we have
\begin{equation*}
R^{\mu}_{\phantom{\mu}\alpha\beta\gamma}+R^{\mu}_{\phantom{\mu}\beta\gamma\alpha}+
R^{\mu}_{\phantom{\mu}\gamma\alpha\beta}=0
\end{equation*}
and also a differential (Bianchi) identity
\begin{equation*}
\bigtriangledown_{\alpha}R^{\mu}_{\phantom{\mu}\nu\beta\gamma}+
\bigtriangledown_{\beta}R^{\mu}_{\phantom{\mu}\nu\gamma\alpha}+
\bigtriangledown_{\gamma}R^{\mu}_{\phantom{\mu}\nu\alpha\beta}=0.
\end{equation*}
Note that for non-symmetric connections, we would have got torsion terms in the right hand sides.

Any model which deals with different metric fields in the (otherwise Riemannian) geometric part and in the matter coupling prescription can be viewed as a model with non-metricity and without torsion. Indeed, being Levi-Civitian for some metric, the connection is definitely symmetric, however it is not metric compatible from the viewpoint of the physical metric which governs the motion of matter. Concrete examples include scalar-tensor models such as $f(R)$ gravity in the Einstein frame.

\subsubsection{Metric compatible connections}

If we have a metric, we can raise and lower the indices. Moreover, if the connection is metric compatible $Q=0$, we can commute the metric with the covariant derivatives. It implies new symmetry properties of the curvature tensor. We can rewrite the commutators of covariant derivatives (\ref{commvec}), (\ref{commform}) as
$$\left[\bigtriangledown_{\mu}\ ,\ \bigtriangledown_{\nu}\right]\xi_{\alpha}=
R_{\alpha\beta\mu\nu}\xi^{\beta}-T^{\beta}_{\hphantom{\beta}\mu\nu}\bigtriangledown_{\beta}\xi_{\alpha},$$
$$\left[\bigtriangledown_{\mu}\ ,\ \bigtriangledown_{\nu}\right]\zeta_{\alpha}=
-R_{\beta\alpha\mu\nu}\zeta^{\beta}-T^{\beta}_{\hphantom{\beta}\mu\nu}\bigtriangledown_{\beta}\zeta_{\alpha}.$$
and deduce:
\begin{equation}
\label{secsymcur}
R_{\alpha\beta\mu\nu}=-R_{\beta\alpha\mu\nu}.
\end{equation}

Note that the antisymmetry (\ref{secsymcur}) is valid also in presence of torsion. Teleparallel gravity belongs to this class of models.

\subsubsection{The Levi-Civita connection}

Finally, if we set both torsion and non-metricity to zero, then our expression for affine connection (\ref{arbconn}) reduces to the Levi-Civita one
$$\Gamma^{\alpha}_{\mu\nu}=\frac12 g^{\alpha\beta}\left(\partial_{\mu}g_{\nu\beta}+\partial_{\nu}g_{\mu\beta}-\partial_{\beta}g_{\mu\nu}\right).$$ In this case we are back to Riemannian geometry, and one can check that our definition of the geodesic line (\ref{geodesic})
coincides with that of the path of the shortest distance between two points (provided they are not too far from each other). Also one can prove another symmetry of the curvature tensor
\begin{equation*}
R_{\mu\nu\alpha\beta}=R_{\alpha\beta\mu\nu}
\end{equation*}
for the  Levi-Civita connection.

As a side note, let us mention that, evidently, the curvature tensor has the maximal amount of symmetry properties when in Riemannian geometry, and therefore it is often much simpler to calculate with the Levi-Civita connection. On the other hand, it is second order (\ref{curvature}) in the connection coefficients, and substituting $\Gamma^{\alpha}_{\mu\nu}=\mathop\Gamma\limits^{(0)}{\vphantom{\Gamma}}^{\alpha}_{\mu\nu}(g)+\delta\Gamma^{\alpha}_{\hphantom{\alpha}\mu\nu}$ where $\mathop\Gamma\limits^{(0)}{\vphantom{\Gamma}}^{\alpha}_{\mu\nu}(g)$ is the Levi-Civita connection for a given metric $g_{\mu\nu}$ and $\delta\Gamma$ is always a tensor, we have a relation
\begin{equation}
\label{varcur}
R^{\alpha}_{\hphantom{\alpha}\beta\mu\nu}(\Gamma)=R^{\alpha}_{\hphantom{\alpha}\beta\mu\nu}(\mathop\Gamma\limits^{(0)})+ \mathop\bigtriangledown\limits^{(0)}{\vphantom{\bigtriangledown}}_{\mu}\delta\Gamma^{\alpha}_{\hphantom{\alpha}\nu\beta}-\mathop\bigtriangledown\limits^{(0)}{\vphantom{\bigtriangledown}}_{\nu}\delta\Gamma^{\alpha}_{\hphantom{\alpha}\mu\beta}
 +\delta\Gamma^{\alpha}_{\hphantom{\alpha}\mu\rho}\cdot\delta\Gamma^{\rho}_{\hphantom{\rho}\nu\beta}-\delta\Gamma^{\alpha}_{\hphantom{\alpha}\nu\rho}\cdot\delta\Gamma^{\rho}_{\hphantom{\rho}\mu\beta}
\end{equation}
which often gives a nicer way of approaching a modified gravity model.

\subsection{Tetrad formulation}

If only for introducing fermions, one needs the tetrad description of general relativity. And so is for teleparallel gravity. We don't need much more from that than is available in any reasonable exposition of standard general relativity in terms of the frame field.

The main essence is that we have a non-degenerate matrix $e^{\mu}_{a}$ composed of components of four vector fields $e_a$, $a=0,1,2,3$ which form an orthonormal ($g_{\mu\nu}e^{\mu}_a e^{\nu}_b=\eta_{ab}$) basis in the tangent space at each point. The inverse matrix is denoted by $e_{\mu}^a$, and the metric components can be calculated as
\begin{equation*}
g_{\mu\nu}=e^a_{\mu}e^b_{\nu}\eta_{ab}, \qquad  g^{\mu\nu}=e_a^{\mu}e_b^{\nu}\eta^{ab}.
\end{equation*}
Very importantly, if we are given a metric, the tetrad fields $e^a_{\mu}$ are only defined up to an arbitrary local Lorentz rotation $e^a_{\mu}(x)\to\Lambda^a_b(x)\cdot e^{b}_{\mu}(x)$ since this is the natural freedom of choosing an orthonormal basis. 

Now, that we have a non-degenerate matrix with indices of two kinds, we can define components of tensor fields with Latin indices by
\begin{equation*}
{\mathcal T}^{a_1,\ldots,a_n}_{\hphantom{a_1,\ldots,a_n}b_1,\ldots,b_m}\equiv e^{a_1}_{\alpha_1}\cdots e^{a_n}_{\alpha_n}{\mathcal T}^{\alpha_1,\ldots,\alpha_n}_{\hphantom{\alpha_1,\ldots,\alpha_n}\beta_1,\ldots,\beta_m} e^{\beta_1}_{b_1}\cdots e^{\beta_m}_{b_m}
\end{equation*}
which amounts to relating tensorial quantities to the chosen frame.

We ought to extend our notion of parallel transport to quantities with the new type of indices. Therefore we introduce some new connection coefficients $\omega^a_{\hphantom{a}\mu b}$ which are usually called spin connection. And now the adopted recipe is that we use $\Gamma$-terms for Greek indices, and $\omega$-terms for Latin indices. For example, $\bigtriangledown_{\mu}T^{a\nu}=\partial_{\mu}T^{a\nu}+\Gamma^{\nu}_{\mu\rho}T^{a\rho}+\omega^a_{\hphantom{a}\mu c}T^{c\nu}$.

Of course,  the spin connection must indeed transform as a connection under the local Lorentz transformations in the space of tetrads: 
\begin{equation}
\label{gaugeom}
e^a_\mu  \longrightarrow \Lambda^a_c e^c_\mu\,, \quad
\omega^a_{\hphantom{a}\mu b} \longrightarrow \Lambda^a_c \omega^c_{\hphantom{a}\mu d}(\Lambda^{-1})^d_b-(\Lambda^{-1})^a_c\partial_{\mu}\Lambda^c_b\, ,
\end{equation}
and with respect to its Latin indices it is assumed to belong to the Lie algebra of the Lorentz group.

If we treat objects with different types of indices as different representations of the same invariant entities, then it is natural to demand that the two notions of parallel transport coincide with each other. Formally it means that the tetrad field (which changes the type of an index) must commute with taking a covariant derivative of a tensor. This is ensured by vanishing of the "full covariant derivative" of the tetrad:
\begin{equation}
\label{fullcovvan}
\partial_{\mu}e^a_{\nu}+\omega^a_{\hphantom{a}\mu b}e^b_{\nu}-\Gamma^{\alpha}_{\mu\nu}e^a_{\alpha}=0
\end{equation}
which is usually imposed when working with tetrads, and which we will also adopt in what follows.

Condition (\ref{fullcovvan}) can be solved straightforwardly to obtain
\begin{equation}
\label{gamom}
\Gamma^{\alpha}_{\mu\nu}=e_a^{\alpha}\left(\partial_{\mu}e^a_{\nu}+\omega^a_{\hphantom{a}\mu b}e^b_{\nu}\right)\equiv e_a^{\alpha}\ {\mathop\mathfrak D}_{\mu}e^a_{\nu}
\end{equation}
with $ {\mathop\mathfrak D}_{\mu}$ being the Lorentz-covariant (with respect to the Latin index only) derivative, or another way around
\begin{equation}
\label{omgam}
\omega^a_{\hphantom{a}\mu b}=e^a_{\alpha}\Gamma^{\alpha}_{\mu\nu}e^{\nu}_b-e^{\nu}_b\partial_{\mu}e^a_{\nu}.
\end{equation}
 In particular, one can find the spin connection $\mathop\omega\limits^{(0)}$ which corresponds to the Levi-Civita connection $\mathop\Gamma\limits^{(0)}(g)$ of a given metric $g$ and obtain the tetrad description of the standard general relativity.

Note that as we have already mentioned, under the condition  (\ref{fullcovvan}) of vanishing full covariant derivative of the tetrad, both $\Gamma^{\alpha}_{\mu\beta}$ and $\omega^a_{\hphantom{a}\mu b}$ represent one and the same connection in different disguises. And indeed, a simple calculation shows that the two curvatures,
\begin{equation*}
R^a_{\hphantom{a}b\mu\nu}(\omega)=\partial_{\mu}\omega^a_{\hphantom{a}\nu b}-\partial_{\nu}\omega^a_{\hphantom{a}\mu b}+\omega^a_{\hphantom{a}\mu c}\omega^c_{\hphantom{c}\nu b}-\omega^a_{\hphantom{a}\nu c}\omega^c_{\hphantom{c}\mu b}
\end{equation*}
and
\begin{equation*}
R^{\alpha}_{\hphantom{\alpha}\beta\mu\nu}(\Gamma)= \partial_{\mu}\Gamma^{\alpha}_{\nu\beta}-\partial_{\nu}\Gamma^{\alpha}_{\mu\beta} +\Gamma^{\alpha}_{\mu\rho}\Gamma^{\rho}_{\nu\beta}-\Gamma^{\alpha}_{\nu\rho}\Gamma^{\rho}_{\mu\beta}\,,
\end{equation*}
are related by a  mere change of the type of indices:
\begin{equation*}
R^{\alpha}_{\hphantom{a}\beta\mu\nu}(\Gamma)=e^{\alpha}_a R^a_{\hphantom{\alpha}b\mu\nu}(\omega) e^b_{\beta}\,.
\end{equation*}

\section{Teleparallel gravity}

Let us now see how we can use different geometries in the tetrad approach.
We immediately note that non-metricity  is automatically zero due to the condition (\ref{fullcovvan}):
\begin{equation*}
\bigtriangledown_{\alpha}g_{\mu\nu}\equiv\eta_{ab}\left(\partial_{\alpha}\left(e^a_{\mu}e^b_{\nu}\right)-\Gamma^{\beta}_{\alpha\mu}e^a_{\beta}e^b_{\nu}-\Gamma^{\beta}_{\alpha\nu}e^a_{\mu}e^b_{\beta}\right)
=-e^b_{\mu}e^c_{\nu}\left(\eta_{ab}\omega^a_{\hphantom{a}\alpha c}+\eta_{ac}\omega^a_{\hphantom{a}\alpha b}\right)=0
\end{equation*}
where in the last step we have taken into account that the matrices $\omega^{\cdot}_{\hphantom{a}\alpha \cdot}$ belong to the Lie algebra of the Lorentz group $SO(1,3)$. 
Therefore we have the antisymmetry (\ref{secsymcur}) property  $R_{\alpha\beta\mu\nu}=-R_{\beta\alpha\mu\nu}$ for the curvature tensor. 

Given the absence of non-metricity, our connection (\ref{arbconn}) is of the form
\begin{equation*}
\Gamma^{\alpha}_{\mu\nu}=\mathop\Gamma\limits^{(0)}{\vphantom{\Gamma}}^{\alpha}_{\mu\nu}(g)+K^{\alpha}_{\hphantom{\alpha}\mu\nu}
\end{equation*}
where $\mathop\Gamma\limits^{(0)}{\vphantom{\Gamma}}^{\alpha}_{\mu\nu}(g)$ is the Levi-Civita connection of the metric $g$, while the tensor
\begin{equation}
\label{contortion}
K_{\alpha\mu\nu}=\frac12\left(T_{\alpha\mu\nu}+T_{\nu\alpha\mu}+T_{\mu\alpha\nu}\right)=\frac12\left(T_{\mu\alpha\nu}+T_{\nu\alpha\mu}-T_{\alpha\nu\mu}\right)
\end{equation}
is known under the name of contortion. 
It is obviously antisymmetric with respect to the two lateral indices:
\begin{equation*}
K_{\alpha\mu\nu}=-K_{\nu\mu\alpha}.
\end{equation*}

In the curvature variation (\ref{varcur}) we can simply substitute $\delta\Gamma$ by $K$ to get
\begin{equation*}
R^{\alpha}_{\hphantom{\alpha}\beta\mu\nu}(\Gamma)=R^{\alpha}_{\hphantom{\alpha}\beta\mu\nu}(\mathop\Gamma\limits^{(0)})+ \mathop\bigtriangledown\limits^{(0)}{\vphantom{\bigtriangledown}}_{\mu}K^{\alpha}_{\hphantom{\alpha}\nu\beta}-\mathop\bigtriangledown\limits^{(0)}{\vphantom{\bigtriangledown}}_{\nu}K^{\alpha}_{\hphantom{\alpha}\mu\beta} +K^{\alpha}_{\hphantom{\alpha}\mu\rho}K^{\rho}_{\hphantom{\rho}\nu\beta}-K^{\alpha}_{\hphantom{\alpha}\nu\rho}K^{\rho}_{\hphantom{\rho}\mu\beta}.
\end{equation*}
After necessary contractions we obtain the fundamental relation for the scalar curvatures:
\begin{equation}
\label{fundrel}
R(\Gamma)=R(\mathop\Gamma\limits^{(0)})+2 \mathop\bigtriangledown\limits^{(0)}{\vphantom{\bigtriangledown}}_{\mu}T^{\mu}+\mathbb T
\end{equation}
where the torsion vector is
\begin{equation*}
T_{\mu}\equiv T^{\alpha}_{\hphantom{\alpha}\mu\alpha}=-T^{\alpha}_{\hphantom{\alpha}\alpha\mu},
\end{equation*}
and the torsion scalar can be written in several equivalent ways:
\begin{eqnarray*}
{\mathbb T} & = &  \frac12 K_{\alpha\beta\mu}T^{\beta\alpha\mu}-T_{\mu}T^{\mu}\\
{} & =  & \frac12 T_{\alpha\beta\mu}S^{\alpha\beta\mu}\\
{} & = &  \frac14 T_{\alpha\beta\mu}T^{\alpha\beta\mu}+\frac12 T_{\alpha\beta\mu}T^{\beta\alpha\mu}-T_{\mu}T^{\mu}
\end{eqnarray*}
with the "superpotential"
\begin{equation}
\label{suppot}
S^{\alpha\mu\nu}\equiv K^{\mu\alpha\nu}+g^{\alpha\mu}T^{\nu}-g^{\alpha\nu}T^{\mu}
\end{equation}
which satisfies the same antisymmetry condition $S^{\alpha\mu\nu}=-S^{\alpha\nu\mu}$ as the torsion tensor itself.
The basic idea of teleparallel gravity would be to set curvature to zero: $R^{\alpha}_{\hphantom{\alpha}\beta\mu\nu}(\Gamma)=0$ and in particular $R(\Gamma)=0$.

\subsection{Teleparallel equivalent of general relativity}

In the classical formulation of teleparallel gravity, one uses the Weitzenb{\" o}ck connection given by 
\begin{equation}
\label{W}
\omega^a_{\hphantom{a}\mu b}=0
\end{equation}
 or equivalently
$\Gamma^{\alpha}_{\mu\nu}= e^{\alpha}_a \partial_{\mu}e^a_{\nu}$
which is obviously curvature-free, $R^{\alpha}_{\hphantom{\alpha}\beta\mu\nu}=0$, and has the following torsion:
$$T^{\alpha}_{\hphantom{\alpha}\mu\nu}=e^{\alpha}_a \left(\partial_{\mu}e^a_{\nu}-\partial_{\mu}e^a_{\nu}\right).$$ 

We can denote the determinant of $e^a_{\mu}$ by $\| e\|=\sqrt{-g}$ and see from the fundamental relation (\ref{fundrel}) that, under the Weitzenb{\" o}ck assumption (\ref{W}), the action
\begin{equation}
\label{telepact}
S=-\int d^4 x \| e\|\cdot {\mathbb T}
\end{equation}
is equivalent to the action $\int d^4 x \sqrt{-g}\cdot R(\mathop\Gamma\limits^{(0)})$ of general relativity modulo a surface term with $2 \mathop\bigtriangledown\limits^{(0)}{\vphantom{\bigtriangledown}}_{\mu}T^{\mu}$.

Note that we discuss pure gravity ignoring the matter fields. There is no problem with minimally coupling bosonic fields since they simply have to interact with the metric $g_{\mu\nu}=e^a_{\mu}e^b_{\nu}\eta_{ab}$ in the usual way. Fermionic fields would require the spin connection $\mathop\omega\limits^{(0)}$ which is admittedly  not so natural in this framework, but it won't be our concern in this paper.

Let us also mention straitaway that the choice (\ref{W}) of $\omega=0$ is locally Lorentz breaking as is evident, if nothing else, from the transformation law (\ref{gaugeom}). This choice allowed us to formulate the model without any explicit use of the spin connection. Sometimes the teleparallel action (\ref{telepact}) with $\omega=0$ is referred to as pure tetrad gravity, as opposed to genuine teleparallel which would have used an arbitrary flat spin connection. 
Note though that this distinction is not common in the literature and very often it may cause a confusion as to what was precisely assumed in a given paper. 

For the teleparallel equivalent of general relativity this issue is not particularly important since the local Lorentz violation is totally contained in the surface term without affecting the equations of motion. However, it will play a role for modified models. We will come to this point soon.

\subsubsection{Equations of motion}

Equations of motion are easily derived for the action (\ref{telepact}). By using the standard method of varying the inverse matrices $\delta e^{\mu}_a  =  -e^{\mu}_b e^{\nu}_a \delta e^b_{\nu}$ and determinants $\delta\|e\|  =  \|e\|\cdot e^{\mu}_a \delta e^a_{\mu}$ we easily get
$$\delta S=-\int d^4 x \|e\|\cdot\left(-2S^{\alpha\mu\nu}T_{\alpha\beta\nu}e^{\beta}_a \delta e^a_{\mu}+{\mathbb T}e^{\mu}_a \delta e^a_{\mu} -2S_{\beta}^{\hphantom{\beta}\mu\alpha}e^{\beta}_a {\mathop\mathfrak D}_{\alpha}\delta e^a_{\mu}\right)$$
where we use the Lorentz-covariant derivative ${\mathop\mathfrak D}$ to account for a non-trivial spin connection if there was one. In the current case ${\mathop\mathfrak D}$  is just a partial derivative since $\omega=0$.

We need to perform integration by parts in the last term which readily results in
$2\delta e^a_{\mu}\cdot \left(\partial_{\alpha}\left(\|e\|\cdot S_{\beta}^{\hphantom{\beta}\mu\alpha}e^{\beta}_a\right)-\|e\|\cdot \omega^b_{\hphantom{b}\alpha a}S_{\beta}^{\hphantom{\beta}\mu\alpha}e^{\beta}_b\right)$. In teleparallel gravity it is often used as it is, in this very non-covariant form. However one can easily do better. Indeed, it is very tempting to make the derivative in the first term into a Levi-Civita-covariant one, $\mathop\bigtriangledown\limits^{(0)}{\vphantom{\bigtriangledown}}_{\nu} S_{a}^{\hphantom{a}\mu\nu}=\frac{1}{\|e\|}\partial_{\nu}\left(\vphantom{\int}\|e\|S_a^{\mu\nu}\right)-\mathop\omega\limits^{(0)}{\vphantom{omega}}^b_{\hphantom{b}\nu a}S_{b}^{\hphantom{b}\mu\nu}$. Correcting for the mismatch of connections with the contortion (\ref{contortion}) tensor $\omega^b_{\hphantom{b}\nu a}-\mathop\omega\limits^{(0)}{\vphantom{omega}}^b_{\hphantom{b}\nu a}=K^b_{\hphantom{b}\nu a}$, one obtains a nice contribution to the equation of motion: $\mathop\bigtriangledown\limits^{(0)}{\vphantom{\bigtriangledown}}_{\nu} S_{a}^{\hphantom{a}\mu\nu}-K^{c}_{\hphantom{c}\nu a}S_{c}^{\hphantom{c}\mu\nu}$.

Finally, using non-degeneracy of tetrads, we get the equation of motion
\begin{equation}
\label{TEGReq}
\mathop\bigtriangledown\limits^{(0)}{\vphantom{\bigtriangledown}}_{\alpha} S_{\beta}^{\hphantom{\beta}\mu\alpha}-S^{\alpha\mu\nu}\left(T_{\alpha\beta\nu}+K_{\alpha\nu\beta}\right)+\frac12 {\mathbb T}\delta^{\mu}_{\beta}=0.
\end{equation}
Is it equivalent to general relativity? Yes! Directly substituting
$$R^{\alpha}_{\hphantom{\alpha}\beta\mu\nu}(\mathop\Gamma\limits^{(0)})=-\left( \mathop\bigtriangledown\limits^{(0)}{\vphantom{\bigtriangledown}}_{\mu}K^{\alpha}_{\hphantom{\alpha}\nu\beta}-\mathop\bigtriangledown\limits^{(0)}{\vphantom{\bigtriangledown}}_{\nu}K^{\alpha}_{\hphantom{\alpha}\mu\beta} +K^{\alpha}_{\hphantom{\alpha}\mu\rho}K^{\rho}_{\hphantom{\rho}\nu\beta}-K^{\alpha}_{\hphantom{\alpha}\nu\rho}K^{\rho}_{\hphantom{\rho}\mu\beta}\right)$$
into the Einstein equation $G^{\mu}_{\beta}(\mathop\Gamma\limits^{(0)},g(e))=0$, one can prove that it is.

\subsection{Possible extensions}

Having found the action  (\ref{telepact}), one is naturally driven to consider possible modifications of gravity in the teleparallel framework. 

One of the first ideas in this direction was the so-called new GR of the Ref. \cite{newGR} which introduces  the Lagrangian density $c_1 T_{\alpha\beta\mu}T^{\alpha\beta\mu}+c_2 T_{\alpha\beta\mu}T^{\beta\alpha\mu}+c_3 T_{\mu}T^{\mu}$ with modified (compared to $\mathbb T$) coefficients $c_i$. On can also add parity violating terms and explore non-linear functions of torsion scalars. And once non-linear functions are introduced, one can also put the divergence  $2 \mathop\bigtriangledown\limits^{(0)}{\vphantom{\bigtriangledown}}_{\mu}T^{\mu}$ inside as an argument \cite{gentel}.

Rather obviously, if the Weitzenb{\" o}ck connection $\omega=0$ is assumed, generically all these modified gravities would violate local Lorentz invariance in the tetrad space, even at the level of equations of motion. This fact used to be a cause of much confusion about probably the most simple modification, the $f(\mathbb T)$ gravity \cite{fT}. Its Lagrangian density is a non-linear function of the torsion scalar from relation (\ref{fundrel}) which is used in the teleparallel equivalent of general relativity. As such it is in a way very similar to $f(R)$ gravity which has one extra degree of freedom due to higher derivatives of the metric in the curvature scalar. The false expectation \cite{fTgood} was that $f(\mathbb T)$ should not increase the number of degrees of freedom since $\mathbb T$  contains only first derivatives of the tetrad.
Of course the catch is that the local Lorentz invariance is broken and this simple count of degrees of freedom is not justified. It has been realised a bit later \cite{tom1, tom2}. 

Another related issue is that Lorentz violating models do care about a particular choice of the tetrad for a given metric. For example, an ansatz for a spherically symmetric solution in spherical coordinates would normally fail for the most natural (diagonal) choice of the tetrad unless a non-trivial flat spin connection is inserted to account for a transition from a Cartesian to a spherical frame.

It even led to introducing the notion of good and bad tetrads \cite{Nicola}. The former represent such ans{\"a}tze which do go through smoothly, while the latter do fail typically requiring $\frac{d^2 f(\mathbb T)}{d{\mathbb T}^2}=0$ which brings us back to the teleparallel equivalent of general relativity with local Lorentz invariance being restored in equations of motion. This is probably a very unique instance in science when failing with a bad ansatz gave rise to a whole new concept.

\section{Review of covariantisation}

As we have already discussed, teleparallel gravity can be naturally covariantised if a spin connection with the proper transformation law (\ref{gaugeom}) is introduced as an additional physical field. Note that it is very important that the spin connection be flat. 

First, otherwise it is not a teleparallel model any longer. 

Second, with an arbitrary spin connection the (exact) variation of the torsion tensor $\delta_{\omega} T^{\alpha}_{\hphantom{\alpha}\mu\nu}=\delta\omega^{\alpha}_{\hphantom{\alpha}\mu\nu}-\delta\omega^{\alpha}_{\hphantom{\alpha}\nu\mu}$ can be used to show that even the simplest action (\ref{telepact}) yields a trivial model instead of general relativity. Indeed, by variation with respect to $\omega$, the equation of motion is $T^{\mu}_{\hphantom{\mu}\alpha\nu}+T_{\nu}\delta_{\alpha}^{\mu}-T_{\alpha}\delta_{\nu}^{\mu}=0$ which (in spacetime dimension $d\neq 2$) entails $T_{\mu}=0$ upon tracing, with the totally trivial final result of
$T^{\mu}_{\hphantom{\mu}\alpha\nu}=0.$

Sometimes the spin connection is treated as not a dynamical field, so that one does not have to make a variation with respect to it. In this case the action with any given choice of the flat spin connection is not invariant, however there is the freedom of making this choice which renders the whole thing invariant again. We don't subscribe to this viewpoint. It seems kind of awkward to have a spacetime-dependent quantity in the action of a fundamental theory which is not subject to variation. Moreover, we feel that such  covariantisation is a very superficial one.

This can be compared to making a model with preferred direction isotropic by explicitly introducing
a vector field in the preferred direction with the correct transformation law for the components under rotations. It is
only our description which gets covariantised in this manner unless the vector, or the spin connection, is subject to an
independent dynamics. A concrete example would be about massive bodies thrown near the surface of Earth. Does it
really put all three directions on equal footing if we covariantly introduce the vector of free fall acceleration with arbitrary
components instead of pointing it along the usual $z$-direction? The real physics is objectively anisotropic at this level, and one would have
to underpin the emergence of the preferred direction in a more fundamental theory.

\subsection{Teleparallel equivalent of general relativity}

Clearly, the proper procedure would be to vary the spin connection constraining it to be flat (curvature-free) or, as some like to call it, purely inertial. Applying a gauge transformation (\ref{gaugeom}) to the zero Weitzenb{\" o}ck connection (\ref{W}), we see that the spin connection we need is of the following form:
\begin{equation}
\label{inerconn}
\omega^a_{\hphantom{a}\mu b}=-(\Lambda^{-1})^a_c \partial_{\mu}\Lambda^c_b
\end{equation}
 where $\Lambda(x)\in SO(1,3)$ is an arbitrary Lorentz matrix. Literally it means that there exists a frame in which $\omega=0$ (Weitzenb{\" o}ck), but a local Lorentz rotation has been done to get away from this frame.

Guided by this idea, we replace the teleparallel (pure tetrad) action (\ref{telepact}) with a slightly more refined version:
\begin{equation}
\label{telepactcov}
S=-\int d^4 x \| e\|\cdot {\mathbb T}(e,\omega(\Lambda))
\end{equation}
in which $e$ and $\Lambda$ should be thought of as independent variables, and the spin connection is given by equation (\ref{inerconn}). Of course, keeping $\Lambda$ fixed but arbitrary would precisely correspond to the "superficial covariantisation" mentioned above.

Varying with respect to $\Lambda$ in the action (\ref{telepactcov}) does not have the disastrous consequences of varying with respect to an arbitrary spin connection. Actually, it does not have any consequences at all. Indeed, from the fundamental relation (\ref{fundrel}) we see that the combination $2 \mathop\bigtriangledown\limits^{(0)}{\vphantom{\bigtriangledown}}_{\mu}T^{\mu}+\mathbb T$ does not depend on $\Lambda$ since $R(\omega(\Lambda))=0$, and therefore the dependence of the action (\ref{telepactcov}) on the spin connection is only a surface term effect \cite{Martin1, MartinEm, GKS}.

In other words,  we have $\delta_{\Lambda}\mathbb T=\delta_{\Lambda} R(\omega)-2\mathop\bigtriangledown\limits^{(0)}{\vphantom{\bigtriangledown}}_{\mu}(\delta_{\Lambda} T^{\mu})$ where $\delta_{\Lambda}$ is given via the usual chain rule $\delta_{\Lambda}(...)=\delta_{\omega}(...)\cdot \delta_{\Lambda}\omega$, and since $R(\omega(\Lambda))\equiv 0$ the variation  is a surface term and does not produce any new equation of motion.
For those who are not satisfied with these simple (and exhaustive) arguments, explicit calculations are given in the  Ref. \cite{GKS}.

\subsubsection{A few remarks}

Note that one could also impose the flatness condition $R^a_{\hphantom{a}b\mu\nu}(\omega)=0$ with a Lagrange multiplier instead of our representation (\ref{inerconn}) having the Lagrangian density of the form
${\mathbb T}(e,\omega)+\lambda_a^{\hphantom{a}b\mu\nu}R^a_{\hphantom{a}b\mu\nu}(\omega)$
where $\lambda_a^{\hphantom{a}b\mu\nu}$ is a Lagrange multiplier with postulated symmetry properties $\lambda^{ab\mu\nu}=-\lambda^{ab\nu\mu}$ and $\lambda^{ab\mu\nu}=-\lambda^{ba\mu\nu}$. 

This is a possible approach \cite{Blago, Nester2} in its own right. Moreover, it can be used in the metric language instead of tetrads, and can also be applied to models of gravity in terms of non-metricity with neither curvature nor torsion \cite{nonmetr}. We think however that for (modified) teleparallel models it might be more instructive to keep the fields $\Lambda$ as the new legitimate players in the game.

Another remark concerns some fundamental considerations about choosing the spin connection. Namely, one might be interested in imposing proper asymptotic conditions for rendering the action finite \cite{Martin1, Martin2} which could be of importance for quantisation of teleparallel gravity. Moreover, sometimes it is claimed that actually there is a spin connection which is naturally associated to a given tetrad \cite{Pereira}.

Such claims refer to a much more elaborated framework, that is to gauging the translations group \cite{newGR, Pereira}. In this approach, due to whatever a reason, one assumes that there exists a global translational symmetry $x^a\to x^a+\epsilon^a$ which can be gauged. Then any tetrad can be separated into two parts, one of which is obtained by a local Lorentz transformation of the "trivial" tetrad $\partial_{\mu}x^a$ thereby prescribing the preferred (flat) spin connection, and another being a non-trivial piece which serves as a gauge potential for the torsion field \cite{Pereira}. This is certainly a more conventional gauge theory compared to our covariantised teleparallel gravity which, having gauged the Lorentz group, strictly keeps it in the purely unphysical sector with vanishing field strength.

It is very unfortunate that there is no accepted terminological distinction to set pure torsion geometry and gauged translations apart. Once we understand that the latter approach starts from the global translations group, it perhaps should no longer come as an unexplainable miracle that people are able to define covariant conservation laws and separate gravity from inertia in this framework \cite{Pereira}, something which is not to get on well with the basic ideas of general relativity.

Probably, the best hopes associated with these constructions hinge upon the idea that the gauge theory approach, which certainly proved so well in doing with other fundamental interactions, might also guide us towards better understanding of gravity \cite{nonmetr, Pereira}, not without a reference to problems of quantisation. However, personally we do not feel that it is time to abandon the unique geometrical flavour of general relativity which makes it so exciting. After all, it's awesome to compute amplitudes of creation from nothing; and all the singularities, Cauchy horizons, causality violations, and time machines might finally be inherent parts of a big beautiful picture.

\subsection{Covariantisation of modified models}

Covariantisation procedures cannot work the same way as above when in modified teleparallel gravities with generic dependence on the tetrad $e^a_{\mu}$ and the torsion $T^{a}_{\hphantom{a}\mu\nu}(e,\omega(\Lambda))$. Since the dependence on the spin connection in generalised models cannot be reduced to a surface term, the variation with respect to the (flat) spin connection does  produce non-trivial equation of motion, though a very benign one. 

One can easily prove that in any covariant teleparallel model this equation is redundant. More precisely, it coincides with the antisymmetric part of the equation of motion for the tetrad field \cite{GKS}. Indeed, by the very definition, the action of a covariant model is identically invariant under combined transformations (\ref{gaugeom}) which for the flat spin connection are obviously in a one-to-one correspondence with variations of the Lorentz matrix $\Lambda$ in its representation (\ref{inerconn}). Therefore, any variation of the action with respect to the flat spin connection (\ref{inerconn}) can be precisely compensated by an appropriate Lorentz rotation of the tetrad. However, the latter is a legitimate variation of the tetrad field which by itself must keep the action stationary, as is independently ensured by the antisymmetric part of the tetrad equation of motion.

In our opinion, this simple argument constitutes a fairly rigorous proof. However, explicit calculations are given in Ref. \cite{GKS} concerning mostly the $f(\mathbb T)$ gravity, and also in Refs. \cite{Martin3, Laur} for more general models.

\subsubsection{Further on $f(\mathbb T)$ gravity}

For the simplest example, in covariant $f(\mathbb T)$ gravity
\begin{equation}
\label{covfT}
S=-\int d^4 x \| e\|\cdot f\left({\mathbb T}(e,\omega(\Lambda)\right)
\end{equation}
one can easily perform all variations and obtain the tetrad equation of motion in the following form \cite{GKS}:
\begin{equation}
\label{fTeqe}
f^{\prime}(\mathbb T)\cdot{\mathop R\limits^{(0)}}{\vphantom{R}}^{\mu\nu}+K^{\mu\nu\alpha}\partial_{\alpha} f^{\prime}(\mathbb T)-T^{\mu}\partial^{\nu} f^{\prime}(\mathbb T)+\frac12 f(\mathbb T)\cdot g^{\mu\nu}=0
\end{equation}
which, due to the local Lorentz breaking in its pure tetrad ($\omega=0$) restriction,  has a non-trivial antisymmetric part 
\begin{equation}
\label{fTeqom}
T^{\alpha\mu\nu}\partial_{\alpha} f^{\prime}(\mathbb T)+T^{\nu}\partial^{\mu} f^{\prime}(\mathbb T)-T^{\mu}\partial^{\nu} f^{\prime}(\mathbb T)=0
\end{equation}
which in turn coincides \cite{GKS} with equation of motion coming from variation with respect to $\Lambda$.

Note that it is this antisymmetric part (\ref{fTeqom}) which normally turns a "bad tetrad" down. It contains only first derivatives of $\omega$ for which reason the flat spin connection of covariant teleparallel gravities is sometimes considered as a non-dynamical variable \cite{Laur}. We must disagree on this point. For example, Hamiltonian  equations are always first order in time derivatives which does not entail that Hamiltonian systems are not truely dynamical. 

This analogy might even go deeper than it seems to, with components of the spin connection being canonically conjugate to each other which would be nicely in line with the $3=\frac62$ new degrees of freedom compared to general relativity.
It is nothing but a mere speculation for now. However, it is evident that we desperately need better understanding of dynamics even in such a simple model as $f(\mathbb T)$. It is not inconceivable that it can be achieved with an accurate treatment of $\Lambda^a_b (x)$ as new fields in the model. 

It would be very good to have a new count of the number of degrees of freedom, more palatable than the truely heroic brute force approach of the Ref. \cite{degrfr} and simultaneously more detailed and reliable than the nice general arguments of the Ref. \cite{Nester1}. The role of the remnant symmetry (those local Lorentz transformations which remain unbroken \cite{remn}) should also be clarified. Last but not least on this very concise list is the conundrum of cosmological perturbations \cite{pertfT}. What is the fundamental reason for the extra modes to be absent from linear cosmological perturbations  around any spatially flat Friedman universe? It is relatively simple to spot an accidental restoration of the local Lorentz invariance in the quadratic (pure tetrad) action for perturbations around Minkowski space, however the general cosmological hide-and-seek looks much more mysterious.

\section{Conclusions}

 Teleparallel gravity provides us with an unprecedented opportunity for constructing new and interesting modified gravity models. And indeed, it is very actively used for cosmological model building \cite{review}. However, as we have seen, the very foundations of the modified teleparallel gravity models do urgently call for better understanding. We hope for major new progress in this direction quite soon.

The Author is greatly indepted to Tomi Koivisto, Jos{\'e} Pereira, Martin Kr{\v s}{\v s}{\'a}k, Yen Chin Ong, and many other people for interesting and fruitful discussions on various topics of teleparallel gravity. Needless to say, this acknowledgement cannot be used to entail that they would agree with any of those views on the subject which have been expressed in this paper.

\end{document}